\documentclass[runningheads]{llncs}
\usepackage[T1]{fontenc}
\usepackage{graphicx}
\usepackage{amssymb,mathtools,array,bbm}
\usepackage[caption=false]{subfig}
\usepackage[ruled]{algorithm2e}
\usepackage{multirow}
\usepackage{booktabs}
\usepackage{wrapfig}
\usepackage{hyperref}

\usepackage{lipsum}
\begin{document}

\newcommand{\ultralmdpp}{\textsc{UltraLMD}\texttt{++}}
\title{Designing a Reliable Lateral Movement Detector Using a Graph Foundation Model}
\titlerunning{Designing a Reliable Lateral Movement Detector Using a GFM}
%
\author{Corentin Larroche}
\authorrunning{C. Larroche}
%
\institute{French Cybersecurity Agency (ANSSI), Paris, France\\
\email{corentin.larroche@ssi.gouv.fr}}
\maketitle              
\begin{abstract}
Foundation models have recently emerged as a new paradigm in machine
learning (ML).
These models are pre-trained on large and diverse datasets and can subsequently
be applied to various downstream tasks with little or no retraining.
This allows people without advanced ML expertise to build ML applications,
accelerating innovation across many fields.
However, the adoption of foundation models in cybersecurity is hindered by
their inability to efficiently process data such as network traffic captures
or binary executables.
The recent introduction of graph foundation models (GFMs) could make a
significant difference, as graphs are well-suited to representing these types
of data.
We study the usability of GFMs in cybersecurity through the lens of one
specific use case, namely lateral movement detection.
Using a pre-trained GFM, we build a detector that reaches state-of-the-art
performance without requiring any training on domain-specific data.
This case study thus provides compelling evidence of the potential of GFMs
for cybersecurity.

\keywords{Intrusion detection \and Anomaly detection \and Graph neural networks.}
\end{abstract}
\section{Introduction}
\label{sec:introduction}

An immediate and lasting impact of large language
models~\cite{brown2020language,touvron2023llama} (LLMs) and their multimodal
counterparts, such as vision-language models~\cite{liu2023visual}, has been the
democratization of machine learning (ML) application development.
Consider for instance natural language processing: whereas building a text
classifier used to require gathering high-quality labeled data, then designing
and training a model, it can now be achieved by simply sending the right prompt
to a pre-existing LLM and extracting information from its answer.
By allowing domain experts without advanced ML skills to build ML
applications, this new paradigm accelerates the adoption of ML in various
domains.

More generally, this trend results from the emergence of
\textbf{foundation models}---that is, models pre-trained on large and diverse
datasets, that can be used for various tasks with little or no retraining.
Beyond LLMs, foundation models have been developed for tasks ranging from time
series forecasting~\cite{ansari2024chronos} to image segmentation and
classification~\cite{kirillov2023segment,oquab2023dino}.
However, their contribution to cybersecurity-related applications has so far
been limited because these applications often deal with peculiar data.
Unlike textual documents, classifying binary executables or network traffic
captures using LLMs is impractical as they differ too much from typical LLM
training data.
On the other hand, graphs are well-suited to representing such data, and
graph neural networks (GNNs) are thus often used in cybersecurity-related ML
applications~\cite{bilot2023graph,bilot2024survey}.

In the last few years, pre-trained GNNs specialized on various graph-related
tasks have started to emerge.
These \textbf{graph foundation models}~\cite{mao2024position} (GFMs) could make
it significantly easier to build ML applications for many cybersecurity-related
use cases, in the same way that LLMs democratized natural language processing.
In previous work~\cite{larroche2024inductive}, we performed an initial
evaluation of GFMs for a specific cybersecurity problem, namely lateral
movement detection.
This task is well-suited to graph-based approaches as it can be framed as
detecting anomalous edges in a
graph~\cite{bowman2020detecting,king2022euler}.
Here, we extend this work into a detailed case study on the following question:
\textbf{
can we reach state-of-the-art performance in lateral movement detection using
a pre-trained GFM, only through careful input graph construction and output
post-processing?
}
By imposing this constraint, we aim to assess the ability of cybersecurity
practitioners without ML expertise to build effective ML applications using
GFMs.
Through experiments on two benchmark datasets, we show that a
GFM-based detector can indeed outperform the state-of-the-art GNN-based
algorithm \textsc{Argus}~\cite{xu2024understanding}.
This demonstrates the potential of GFMs in cybersecurity-related use cases,
opening the way for further research on other cybersecurity problems involving
graph-structured data.

In summary, we make the following contributions:
\begin{itemize}
	\item We extend previous work on GFM-based lateral movement detection,
		improving the design of the detector with an input graph construction
		module that retrieves the most relevant contextual data and an
		output post-processing module that leverages domain knowledge on
		lateral movement to eliminate false positives.
	\item We thoroughly demonstrate the impact of these improvements
		on detection performance through experiments on two benchmark datasets.
	\item Overall, we provide evidence that GFMs can already be used in
		valuable cybersecurity-related applications, motivating further
		research on this topic.
\end{itemize}

The rest of this paper is structured as follows.
We start with necessary background on lateral movement detection and graph
foundation models in Section~\ref{sec:background}.
Then, we describe the design of our GFM-based detector in
Section~\ref{sec:design}, and report on our experiments in
Section~\ref{sec:experiments}.
Finally, we discuss the limitations and possible extensions of our work in
Section~\ref{sec:discussion}.

\section{Background}
\label{sec:background}

We first provide some necessary background and definitions.
Section~\ref{sec:background:problem} defines lateral movement detection,
then Section~\ref{sec:background:related} reviews past research on this
problem.
Finally, we provide an introduction to graph foundation models in
Section~\ref{sec:background:graph}, focusing on
\textsc{Ultra}~\cite{galkin2024towards}, which is the model we use in this
work.

\subsection{Problem Statement --- Lateral Movement Detection}
\label{sec:background:problem}

\textbf{Lateral movement} is a widespread offensive tactic that consists in
propagating from one compromised host to another within the target network.
It is a critical phase of advanced, multi-step intrusions as it allows the
attacker to explore the network and move closer to their end goal (e.g.,
stealing sensitive data from a given server or deploying ransomware on all
hosts).
Many techniques and procedures exist for lateral movement, often leveraging
legitimate tools and credentials~\cite{attack8}.
This diversity makes detecting lateral movement challenging.

Internal network traffic and authentication logs are useful data sources in
the search for lateral movements\footnote{Note that our threat model assumes
the integrity of these data sources.}.
Such data can be efficiently represented as a graph whose nodes are the hosts
of an enterprise network, with edges standing for information flows between
hosts.
Additional nodes representing user accounts can also be included in this graph
when authentication logs are available.
The graph is typically directed, and it can also be made temporal (network
flows and authentication events are timestamped) and heterogeneous (there are
several edge types representing, for instance, network protocols or
authentication packages).
Lateral movement then results in unexpected edges in this graph.
From a mathematical perspective, lateral movement detection can thus be phrased
as a \textbf{graph-based anomaly detection} problem.

\paragraph{Key definitions and notations.}
We consider a \textbf{graph sequence}
$\big(\mathcal{G}_t\big)_{t\geq{0}}$, where each graph
$\mathcal{G}_t=(\mathcal{V}_t,\mathcal{E}_t,\mathcal{R}_t)$ represents events
recorded in a time window $t$ and is defined by its node set $\mathcal{V}_t$,
edge set $\mathcal{E}_t$ and relation (or edge type) set $\mathcal{R}_t$.
Edges, defined by a source node $u$, an edge type $r$, and a destination node
$v$, are denoted
$(u,r,v)\in\mathcal{V}_t\times\mathcal{R}_t\times\mathcal{V}_t$.
Note that such heterogeneous graphs are typically called
\textbf{knowledge graphs} in the literature.
The goal of lateral movement detection is to build an
\textbf{anomaly scoring function}
$s$ such that $s(u,r,v;t)$ is large if the edge $(u,r,v)$ is anomalous in the
graph $\mathcal{G}_t$.
Finally, we define the \textbf{union} of two graphs as the graph whose node,
edge, and relation sets are the union of theirs.

\subsection{Related Work --- From Heuristics to Graph Neural Networks}
\label{sec:background:related}

We now review the main approaches to lateral movement detection in the
literature, distinguishing three main categories: heuristics, linear and
multilinear link predictors, and graph neural networks.

\paragraph{Heuristics and pattern mining.}
Early work on lateral movement detection uses rather simple models and
heuristics.
In particular, a straightforward way to define anomalous edges in a temporal
graph is to count the occurrences of each edge over time.
The \textbf{edges that appear the least frequently} are then deemed anomalous.
Additional edge features, such as edge type or timestamp, can also be taken
into account when looking for rare edges~\cite{bowman2024nethawk}.
The main shortcoming of this basic heuristic is its tendency to generate many
false positives, as rare edges are often observed due to benign errors or
legitimate behaviors.
A typical way to mitigate this problem is to look for \textbf{clusters
of rare edges} sharing a common source
node~\cite{neil2013scan,neil2013towards,%
turcotte2014detecting,bowman2024nethawk}, since attackers often use one
compromised host as a stepping stone to explore the rest of the network.
Rare connections between hosts can also be correlated with other possible
signs of lateral movement, such as the creation of a new service on the target
host~\cite{liu2018latte}.
Alternatively, simple statistical models can be used to characterize normal
rare edges, thus weeding out false positives~\cite{ho2021hopper}, or to detect
anomalous paths in the host communication graph~\cite{bohara2017unsupervised}.
Finally, a closely related approach consists in mining frequent patterns among
edges instead of frequent edges~\cite{siadati2017detecting}.
These patterns then also cover rarely observed edges that share some
distinctive traits with frequent edges, ensuring that such rare but legitimate
edges are not mistakenly flagged as malicious.

\paragraph{Linear and multilinear models.}
Detecting anomalous edges in a graph is in fact complementary to the
well-studied problem of \textbf{link prediction}, i.e., predicting which new
edges are
the most likely to appear in a given graph.
Since anomalous edges are those least likely to appear, a good link predictor
can also be used as an anomaly detector.
Several models that were initially designed for link prediction, typically for
recommender systems, were thus subsequently used for lateral movement
detection.
These models include matrix factorization~\cite{passino2022graph}, tensor
factorization~\cite{eren2020multi,eren2023general}, factorization
machines~\cite{tang2017reducing} and latent space models~\cite{lee2022anomaly},
which are all multilinear models.
More specifically, these models compute a latent vector for each node during
training, then define the probability of an edge as a function of the dot
product between its endpoints' latent vectors.
A closely related approach relies on graph embedding algorithms such as
node2vec~\cite{grover2016node2vec}, which also compute a latent vector (also
called embedding) for each node.
Linear models can then take these vectors as input in order to predict normal
edges and/or detect anomalous ones~\cite{wei2019age,zhao2019ctlmd,%
bowman2020detecting,paudel2022pikachu}.

\paragraph{Graph neural networks.}
Finally, more recent and sophisticated link prediction algorithms go beyond
linear models and use multilayer graph neural networks~\cite{corso2024graph}.
As a consequence, research on lateral movement detection has also adopted
GNNs.
Early contributions treated internal network traffic or authentication logs
as a static graph, training standard GNNs on all past data to predict future
edges~\cite{liu2020mltracer,sun2022hetglm}.
Subsequent work introduced a \textbf{temporal perspective} by representing
connections between internal hosts as a sequence of graphs.
This sequence can then be modeled by intertwining a GNN and a recurrent
neural network (RNN), giving the model the ability to detect temporal anomalies
in addition to structural ones~\cite{king2022euler,khoury2024jbeil,%
xu2024understanding}.
The current state-of-the-art lateral movement detection algorithm,
\textsc{Argus}~\cite{xu2024understanding}, relies on this RNN-GNN combination.

\subsection{Graph Foundation Models for Link Prediction}
\label{sec:background:graph}

Following the emergence of powerful foundation models for modalities such as
text and images, the graph machine learning community has started exploring
the possibility of building such foundation models for graph-related tasks.
As a result, such models have been proposed for various tasks, including node
classification~\cite{huang2023prodigy,zhao2024graphany}, node anomaly
detection~\cite{qiao2025anomalygfm}, and link
prediction~\cite{dong2024universal,shen2024zero}.
In particular, \textbf{link prediction in knowledge graphs} has been addressed
by several GFMs~\cite{galkin2024towards,zhang2024trix}.
This is especially interesting for lateral movement detection, which can be
framed as detecting anomalous edges in a knowledge graph.

This work focuses on \textsc{Ultra}~\cite{galkin2024towards}, a GFM designed
for link prediction in knowledge graphs.
\textsc{Ultra} builds upon the Neural Bellman-Ford Network~\cite{zhu2021neural}
(NBFNet) architecture to predict possible edges in any
knowledge graph $\mathcal{G}=(\mathcal{V},\mathcal{E},\mathcal{R})$ through
the following steps.
First, reciprocal edges are added to the graph $\mathcal{G}$: for each relation
$r\in\mathcal{R}$, a reciprocal relation $r^{-1}$ is defined and a new edge
$(v,r^{-1},u)$ is created for each existing edge $(u,r,v)\in\mathcal{E}$.
Then, a graph of relations \(
	\mathcal{G}_{\mathrm{R}}=(
		\mathcal{R},\mathcal{E}_{\mathrm{R}},
		\mathcal{R}_{\mathrm{R}}
	)
\) is built.
Its nodes are the relations of the original knowledge graph $\mathcal{G}$, and
an edge from relation $r$ to relation $r'$ indicates that there exist two
edges in $\mathcal{G}$, with respective types $r$ and $r'$, that have at least
one node in common.
This common node can be either the source or destination of each of the two
edges, thus there are four possible types of interactions between relations
$r$ and $r'$ (i.e., $|\mathcal{R}_{\mathrm{R}}|=4$).
A first NBFNet is trained to compute embeddings of the relations in
$\mathcal{R}$ given the graph of relations $\mathcal{G}_{\mathrm{R}}$.
Then, given the original graph $\mathcal{G}$ and these relation embeddings, a
second NBFNet predicts the probability of any edge
$(u,r,v)\in\mathcal{V}\times\mathcal{R}\times\mathcal{V}$.
The authors of \textsc{Ultra} \textbf{pre-trained} three such models on a set
of knowledge graphs representing knowledge bases from several domains.
They showed that these models achieved competitive link prediction performance
on knowledge graphs not included in their training set, which makes them
foundation models for link prediction in knowledge graphs.

From a high-level perspective, a pre-trained \textsc{Ultra} model is a
function that takes as input a knowledge graph
$\mathcal{G}=(\mathcal{V},\mathcal{E},\mathcal{R})$ (which we call the
\textbf{context graph}) and a triple
$(u,r,v)\in\mathcal{V}\times\mathcal{R}\times\mathcal{V}$, and returns a score
$g(u,r,v;\mathcal{G})$ that is large if the edge $(u,r,v)$ is likely to
appear in the graph $\mathcal{G}$.
In previous work~\cite{larroche2024inductive}, we showed that the pre-trained
models introduced in the \textsc{Ultra} paper~\cite{galkin2024towards} could be
used for lateral movement detection without retraining, achieving competitive
detection performance with respect to standard multilinear models.
Here, we extend this work by designing a better detector using the
\textsc{Ultra50g} model.
This GFM has 177K parameters and is openly available on
GitHub\footnote{\url{https://github.com/DeepGraphLearning/ULTRA}} and
HuggingFace\footnote{\url{https://huggingface.co/mgalkin/ultra_50g}}.
Our hypothesis is that significant performance gains can be unlocked using the
same pre-trained model, only through changes in context graph construction and
output post-processing---an approach inspired by LLM-based applications.

\section{Design of Our Lateral Movement Detector}
\label{sec:design}

We improve upon our initial algorithm,
\textsc{UltraLMD}~\cite{larroche2024inductive}, by drawing inspiration from
the state-of-the-art GNN-based detector
\textsc{Argus}~\cite{xu2024understanding}.
Through this approach, we investigate the possibility of replicating the
performance of a specially designed GNN by cleverly querying a
pre-trained GFM.
We provide an overview of our new detector, called \ultralmdpp, in
Section~\ref{sec:design:overview},
then describe its main components in further detail: context graph construction
(Section~\ref{sec:design:input}), anomaly scoring
(Section~\ref{sec:design:edge}), and output post-processing
(Section~\ref{sec:design:output}).

\subsection{Overview}
\label{sec:design:overview}

\begin{figure}[t]
	\centering
	\includegraphics[width=\textwidth]{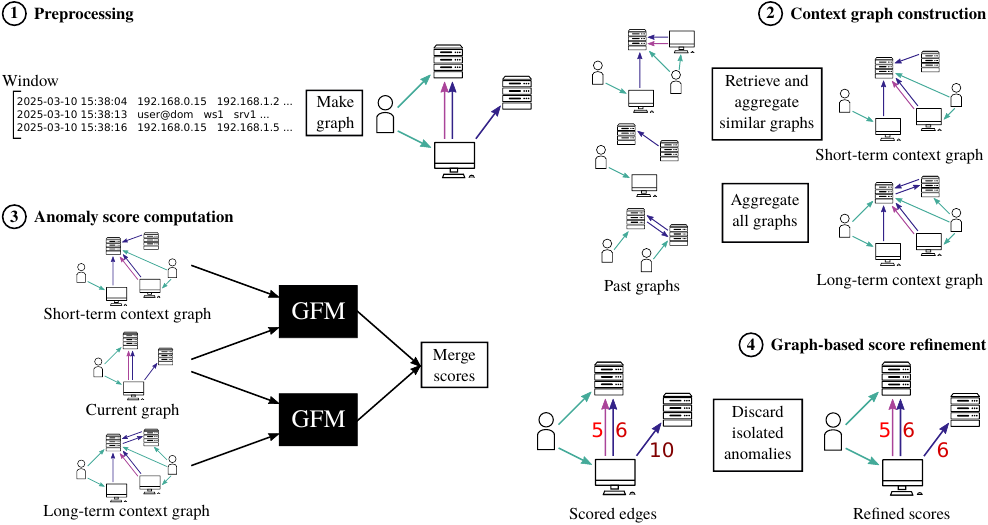}
	\caption{High-level workflow of \ultralmdpp.}
	\label{fig:workflow}
\end{figure}

Given a sequence of authentication events and/or network flows, our lateral
movement detector applies the following steps, illustrated in
Figure~\ref{fig:workflow}.
First, the events are split into \textbf{fixed-length windows}.
Then, for each time window $t$, we build a graph $\mathcal{G}_t$ representing
the events within this window.
In order to compute anomaly scores for the edges in $\mathcal{G}_t$, we first
need to build a \textbf{context graph} $\mathcal{H}_t$ using
\textbf{past events} (Section~\ref{sec:design:input}).
This context graph can then be passed as input to the GFM, which predicts
how likely the edges in $\mathcal{G}_t$ are given the context graph
$\mathcal{H}_t$.
\textbf{Anomaly scores} are derived from these predictions
(Section~\ref{sec:design:edge}).
Finally, these anomaly scores are \textbf{refined} by leveraging the
\textbf{structure of the graph}
$\mathcal{G}_t$, building upon the idea that lateral movement edges tend
to be clustered into connected regions of the network
(Section~\ref{sec:design:output}).

\subsection{Context Graph Construction}
\label{sec:design:input}

At each time step $t$, we observe a new graph
$\mathcal{G}_t=(\mathcal{V}_t,\mathcal{E}_t,\mathcal{R}_t)$.
Our goal is to compute an anomaly score for each edge in $\mathcal{E}_t$ by
comparing the graph $\mathcal{G}_t$ with past graphs
$\left\{\mathcal{G}_{t'};\,t'<t\right\}$.
\textsc{Argus}~\cite{xu2024understanding} adopts a temporal modeling approach
to incorporate the fact that \textbf{a given edge might
be normal at some time steps and anomalous at others} (consider for instance
a remote authentication to a server at 2p.m. on a Monday versus at 11p.m. on
a Saturday).
To that end, it uses a combined RNN-GNN model to encode the sequence of
past graphs and make time-dependent predictions for the probability of each
edge in $\mathcal{E}_t$.
Since we use a GFM to score edges, we cannot modify its architecture to include
an RNN.
We thus adopt a different approach, which focuses on the context graph used by
the GFM to predict new edges.

As explained in Section~\ref{sec:background:graph}, we consider a GFM that
takes as input a context graph and a possible edge, and predicts the
probability of that edge appearing in the context graph.
\textbf{Time-dependent information} can thus be passed to the GFM through the
\textbf{construction of this context graph}.
Specifically, given the current graph $\mathcal{G}_t$ and the sequence of past
graphs $\left\{\mathcal{G}_{t'};\,t'<t\right\}$, we build two context graphs
$\mathcal{H}_t^{\mathrm{L}}$ and $\mathcal{H}_t^{\mathrm{S}}$ encoding
long-term and short-term context, respectively.
The long-term context graph is simply the union of all past graphs, \(
	\mathcal{H}_t^{\mathrm{L}}=\bigcup_{t'<t}\mathcal{G}_{t'}.
\)
As for the short-term context graph $\mathcal{H}_t^{\mathrm{S}}$, we build it
using a method analogous to retrieval-augmented generation
(RAG~\cite{lewis2020retrieval}) for LLMs.
Given a \textbf{similarity function} $c$ that compares two graphs, we compute
the similarity between the current graph $\mathcal{G}_t$ and each past graph.
Denoting $\mathcal{C}$ the set of indices corresponding to the $K$ past graphs
most similar to $\mathcal{G}_t$ (for some fixed integer $K$), the short-term
context graph is then defined as \(
	\mathcal{H}_t^{\mathrm{S}}=\bigcup_{t'\in\mathcal{C}}\mathcal{G}_{t'}.
\)

The similarity function $c$ we use in our experiments is simple and
computationally inexpensive.
It is defined as the sum of two components, $c=c_{\mathrm{V}}+c_{\mathrm{E}}$,
where $c_{\mathrm{V}}$ is the Jaccard index of the respective node sets of the
two graphs, \[
	c_{\mathrm{V}}(\mathcal{G}_t,\mathcal{G}_{t'})=\frac{
		|\mathcal{V}_t\cap\mathcal{V}_{t'}|
	}{
		|\mathcal{V}_t\cup\mathcal{V}_{t'}|
	},
\] and $c_{\mathrm{E}}$ measures the similarity of the two graphs in terms of
edge types.
Let $\theta_r(\mathcal{G}_t)=|\{(u,r',v)\in\mathcal{E}_t:\,r'=r\}|$ be the
number of edges of type $r$ in the graph $\mathcal{G}_t$.
Then the relation similarity function $c_{\mathrm{E}}$ is defined as \[
	c_{\mathrm{E}}(\mathcal{G}_t,\mathcal{G}_{t'})=\frac{
		\tilde{\boldsymbol{\theta}}(\mathcal{G}_t)
		\cdot
		\tilde{\boldsymbol{\theta}}(\mathcal{G}_{t'})
	}{
		\left\|\tilde{\boldsymbol{\theta}}(\mathcal{G}_t)\right\|
		\left\|\tilde{\boldsymbol{\theta}}(\mathcal{G}_{t'})\right\|
	},
	\quad
	\tilde{\boldsymbol{\theta}}(\mathcal{G})=\left(\frac{
		\theta_r(\mathcal{G})-\bar{\theta}_r
	}{
		\sigma_r
	}\right)_{r\in\mathcal{R}_t\cup\mathcal{R}_{t'}},
\] where $\bar{\theta}_r$ and $\sigma_r$ denote the mean and standard deviation
of $\theta_r$ over the set of graphs
$\left\{\mathcal{G}_{t'};\,t'\leq{t}\right\}$, respectively.
In other words, $c_{\mathrm{E}}$ is the cosine similarity between the
standardized vectors of relation counts of the two graphs.

\subsection{Anomaly Score Computation}
\label{sec:design:edge}

Given the current graph $\mathcal{G}_t$ and the short and long-term context
graphs $\mathcal{H}_t^{\mathrm{S}}$ and $\mathcal{H}_t^{\mathrm{L}}$, we can
now use the GFM to compute anomaly scores for the edges of $\mathcal{G}_t$.
As stated in Section~\ref{sec:background:graph}, the GFM is a function that
computes a score $g(u,r,v;\mathcal{H})$ given a context graph
\(
	\mathcal{H}=(
		\mathcal{V}_{\mathcal{H}},
		\mathcal{E}_{\mathcal{H}},
		\mathcal{R}_{\mathcal{H}}
	)
\) and an edge \(
	(u,r,v)\in\mathcal{V}_{\mathcal{H}}\times\mathcal{R}_{\mathcal{H}}
	\times\mathcal{V}_{\mathcal{H}}
\), with higher scores meaning more likely edges.
We turn these scores into anomaly scores as follows.
First, we define the \textbf{predicted probability} of the destination node
$v$ given
the source $u$, the edge type $r$, and the context graph $\mathcal{H}$ as \[
	p(v\mid u,r,\mathcal{H})=\frac{
		\exp\left(g(u,r,v;\mathcal{H})\right)
	}{
		\sum_{v'\in\mathcal{V}_{\mathcal{H}}}
			\exp\left(g(u,r,v';\mathcal{H})\right)
	}.
\] The probability of the source node $u$ given the destination $v$, the edge
type $r$, and the context graph $\mathcal{H}$ can be defined in a similar way
using the reciprocal relation $r^{-1}$, defined in
Section~\ref{sec:background:graph}.
The \textbf{anomaly score} of the edge $(u,r,v)$ given the context graph
$\mathcal{H}$
is then defined as \[
	s(u,r,v;\mathcal{H})=-\log\left(
		\min\left\{
			p(v\mid u,r,\mathcal{H}),
			p(u\mid v,r^{-1},\mathcal{H})
		\right\}
	\right).
\]

Two specific cases are handled differently.
First, if $(u,r,v)\in\mathcal{E}_{\mathcal{H}}$, then we set
$s(u,r,v;\mathcal{H})=0$.
In other words, \textbf{known edges are considered normal}.
Note that while the long-term context graph contains all previously observed
edges, the short-term context graph only contains edges from the past graphs
most similar to the current graph.
As a consequence, an edge that was previously observed in a different context
can still be declared anomalous in the present context by our detector.
The other specific case arises when scoring an edge whose
\textbf{type does not exist}
in the context graph, that is, $r\notin\mathcal{R}_{\mathcal{H}}$.
Since the GFM is unable to compute scores for such unknown edge types, we then
define the anomaly score as the average over the set of known edge types
$\mathcal{R}_{\mathcal{H}}$.
With these definitions, the anomaly score for an edge $(u,r,v)$ given the
short and long-term context graphs $\mathcal{H}_t^{\mathrm{S}}$ and
$\mathcal{H}_t^{\mathrm{L}}$ is
\begin{equation}
	s(u,r,v;t) = \frac{1}{2}\left(
		s(u,r,v;\mathcal{H}_t^{\mathrm{S}})
		+ s(u,r,v;\mathcal{H}_t^{\mathrm{L}})
	\right).
	\label{eq:score}
\end{equation}

\subsection{Graph-Based Score Refinement}
\label{sec:design:output}

In addition to individually scoring each edge in the current graph,
looking for \textbf{connected clusters of anomalous edges} is a widespread
approach in the lateral movement detection literature~\cite{neil2013scan,%
bowman2020detecting,lee2022anomaly,bowman2024nethawk}.
The motivation for this strategy is that attackers typically perform several
lateral movements originating from the same compromised host, whereas rare
but legitimate activity is randomly scattered across the internal network.
\textsc{Argus} leverages this domain knowledge through the design of its
decoder (i.e., the last GNN layer that predicts likely edges).
Once again, since we build upon a pre-trained model, we cannot change
its architecture and must therefore look for a different method.

We draw inspiration from Bowman et al.~\cite{bowman2020detecting}, who
proposed \textbf{discarding all anomalous edges that do not share a node with
at least one other anomalous edge}.
This idea can be translated into the following edge score refinement method,
formally described in Algorithm~\ref{alg:refinement}
of Appendix~\ref{sec:graph}.
For each edge $e$ in the current graph $\mathcal{G}_t$, we retrieve the
anomaly scores (as defined in Equation~\ref{eq:score}) of all edges in
$\mathcal{G}_t$ that share at least one node with $e$.
Then, there are two possibilities: either there is at least one edge in this
set with a higher anomaly score than $e$, thus for any detection threshold low
enough to declare $e$ anomalous, $e$ will be adjacent to another anomalous
edge; or $e$ is the highest-scored edge within its neighborhood.
In that second case, we set the score of $e$ to the maximum of the scores of
adjacent edges, so that $e$ is only declared anomalous when there is at least
one other anomalous edge within its neighborhood.
In practice, this score refinement method can be implemented efficiently using
a message passing algorithm.

\section{Experiments}
\label{sec:experiments}

We now empirically study the behavior and performance of \ultralmdpp,
specifically investigating the following questions.
First, since our goal is to build a lateral movement detector that is
at least as reliable as the state-of-the-art algorithm
\textsc{Argus}~\cite{xu2024understanding}, we compare their respective
\textbf{detection performance} on two public benchmark datasets.
Secondly, since \ultralmdpp\ differs from the original \textsc{UltraLMD}
detector in two aspects (namely, context graph retrieval and score refinement),
we are interested in evaluating the \textbf{exact contribution of each of
these two improvements}.
Third, an expected benefit of using a GFM-based lateral movement detector
is that it can handle concept drift without retraining.
Thus we investigate the temporal dynamics of the distribution of anomaly scores
to assess the \textbf{robustness of \ultralmdpp\ to concept drift}.
Finally, the \textbf{computational cost} of \ultralmdpp\ is an
important aspect of its real-world usability.
We thus study the run time of each of its components (i.e.,
context graph construction, score computation, and score refinement).

We implemented \ultralmdpp\ in Python and make the code openly
available\footnote{\url{https://github.com/cl-anssi/UltraLMD}}.
As for \textsc{Argus}, we used the implementation provided by the
authors\footnote{\url{https://github.com/C0ldstudy/Argus}}.
We run our experiments on a 2.2GHz, 40-core CPU with 384GB of RAM, and an
Nvidia GeForce RTX 2080 Ti GPU with 11GB of memory.
The datasets and evaluation metrics used in our experiments are described in
Section~\ref{sec:experiments:datasets}.
We then present our results on the two datasets in
Sections~\ref{sec:experiments:optc} and~\ref{sec:experiments:lanl},
respectively.

\subsection{Datasets and Metrics}
\label{sec:experiments:datasets}

We compare \ultralmdpp\ and \textsc{Argus} on two benchmark
datasets: the "Operationally Transparent Cyber" (\textbf{OpTC}) dataset
released by DARPA in 2020, and the "Comprehensive, Multi-Source
Cyber-Security Events" (\textbf{LANL})
dataset released in 2015 by the Los Alamos National
Laboratory~\cite{kent2015authentication,kent2015cyberdata}.
OpTC consists of host and network logs generated over nine days in a
simulated enterprise network.
Three distinct attacks were carried out by pentesters against this simulated
network, two of which comprise a lateral movement phase.
The first six days are used for training \textsc{Argus}, and both
\textsc{Argus} and \ultralmdpp\ are evaluated on the last
three days, during which the attacks take place.
As for the LANL dataset, it contains anonymized host logs and network flows
collected over 58 days within LANL's enterprise network.
A red team exercise took place during this time frame, and lateral movement
events are labeled.
Since the first of these events occurs during the $42^{\mathrm{nd}}$ hour,
we use the first 41 hours for training \textsc{Argus} and the remaining data
for evaluation.
Note that while the authors of \textsc{Argus} used the same datasets in their
experiments, we perform different preprocessing steps here,
which leads to discrepancies with respect to the results reported in the
\textsc{Argus} paper~\cite{xu2024understanding}.
See Appendix~\ref{sec:additional} for more details.

{\setlength{\tabcolsep}{.5em}
\begin{table}[t]
	\caption{Datasets used in our experiments. LM stands for lateral movement
	edges; Min./Med./Max. is the minimum/median/maximum number of
	edges per time step.}
	\centering
	\begin{tabular}{lrrrrrrrr}
		\toprule
			\multirow{2}*{\textbf{Dataset}}
			& \multirow{2}*{\textbf{Hosts}}
			& \multirow{2}*{\textbf{Users}}
			& \textbf{Time}
			& \multicolumn{5}{c}{\textbf{Edges}} \\
			& & & \textbf{steps}
			& \textbf{Min.} & \textbf{Med.} & \textbf{Max.}
			& \textbf{Types} & \textbf{LM} \\
		\midrule
			OpTC & 898 & 1,505 & 2,113 & 0 & 5,795 & 17,782 & 39 & 626 \\
			LANL & 16,377 & 25,701 & 1,392 & 72,717 & 151,343.5
			& 286,350 & 106 & 500 \\
		\bottomrule
	\end{tabular}
	\label{tab:description}
\end{table}
}

\begin{figure}[t]
	\centering
	\includegraphics[width=\textwidth]{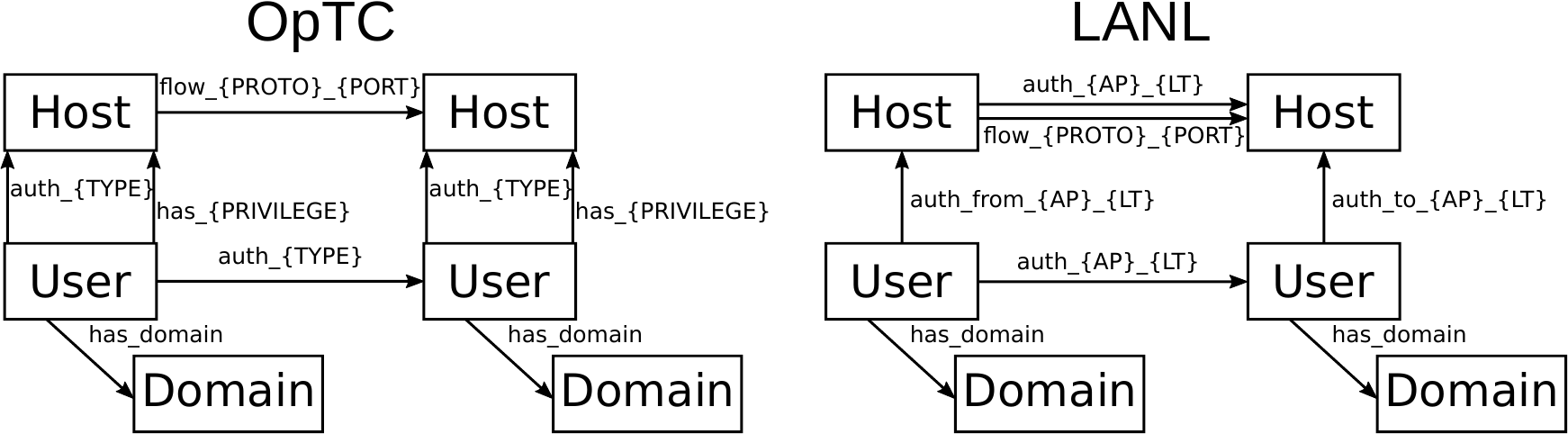}
	\caption{Knowledge graph representation of the two datasets for
	\ultralmdpp.
	AP and LT stand for authentication package and logon type,
	respectively.
	Nodes and edges representing node types are not displayed for the
	sake of readability.}
	\label{fig:schema}
\end{figure}

\paragraph{Graph construction.}
Both \textsc{Argus} and \ultralmdpp\ require representing the datasets
as sequences of graphs.
First of all, we chunk both datasets into fixed-length time windows.
We use the same window lengths as the \textsc{Argus} paper, namely one hour
for the LANL dataset and six minutes for the OpTC dataset.
For \textsc{Argus}, we then use the same graph construction method as the
original paper.
As for \ultralmdpp, the graphs are built as follows.
For each remote authentication event, we create an edge from the source host
to the destination host.
The type of the edge is the pair (authentication package, logon type).
We also create host-host edges for network flow events, whose type is the
pair (transport protocol, destination port).
To limit the number of edge types, only the 20 flow types most frequently seen
during the training period are kept, and all
other types are replaced with a single "Other" type.
We add user-host edges for each authentication event, which
indicate that the user has logged on from or to the host.
Finally, we define five node types: user account, computer account, built-in
account (e.g., SYSTEM or LOCAL SERVICE), computer, and domain.
These types are represented as nodes and linked to other nodes with edges of
a special type.
See Figure~\ref{fig:schema} for a summary of these graph representations,
and Table~\ref{tab:description} for a summary of the datasets.
Note that for consistency with \textsc{Argus}, only host-host edges
representing network flows in the OpTC dataset and remote authentications in
the LANL dataset are given anomaly scores.

\paragraph{Evaluation metrics.}
We compare the detection performance of \textsc{Argus} and \ultralmdpp\ using
the following metrics.
First, the area under the ROC curve (\textbf{AUC})
is a standard metric for binary classification, which captures the trade-off
between the true positive rate (TPR, also called recall) and the false positive
rate (FPR) at various detection thresholds.
Since the AUC tends to be overly optimistic in highly imbalanced settings
(base rate fallacy~\cite{arp2022and}), we
also compute the average precision (\textbf{AP}), which evaluates the ratio
between the number of true and false positives at different true positive
rates.
This metric puts more emphasis on low-FPR settings, which corresponds to the
operational constraints of intrusion detection.
Finally, to get a more realistic picture of each detector's reliability in a
real-world deployment, we compute the recall for various detection budgets $B$
(\textbf{Rec@}$\boldsymbol{B}$).
In other words, Rec@$B$ is the proportion of lateral movement edges that
are detected when the top $B$ most anomalous edges in each time window are
investigated.

\subsection{Results on OpTC}
\label{sec:experiments:optc}

We run both \textsc{Argus} and \ultralmdpp\ on the OpTC dataset, using the same
hyperparameters as the authors for \textsc{Argus} and setting the number $K$
of past graphs to include in the short-term context of \ultralmdpp\ to 100.
The results of these experiments are shown in
Table~\ref{tab:res_optc}.
We now discuss them in further detail.

{\setlength{\tabcolsep}{.3em}
\begin{table}[t]
	\caption{Results of our experiments on OpTC.
	}
	\centering
	\begin{tabular}{lrrrrr}
		\toprule
		\textbf{Algorithm}
		& \textbf{AUC} & \textbf{AP}
		& \textbf{Rec@}$\boldsymbol{3}$
		& \textbf{Rec@}$\boldsymbol{5}$
		& \textbf{Rec@}$\boldsymbol{10}$\\
		\midrule
		\textsc{Argus}~\cite{xu2024understanding}
		& .9300 & .0240 & .0064 & .0113 & .0273 \\
		\midrule
		\textsc{UltraLMD}~\cite{larroche2024inductive}
		& .9826 & .1003 & .0208 & .0319 & .0559 \\
		\textsc{UltraLMD} + retrieval
		& \textbf{.9909} & .1506 & \textbf{.0256} & .0367 & .0607 \\
		\textsc{UltraLMD} + refinement
		& .9826 & .1003 & .0192 & .0319 & .0559 \\
		\midrule
		\ultralmdpp
		& \textbf{.9909} & \textbf{.1510} & \textbf{.0256}
		& \textbf{.0383} & \textbf{.0623} \\
		\bottomrule
	\end{tabular}
	\label{tab:res_optc}
\end{table}
}

First of all, \textbf{\ultralmdpp\ consistently outperforms
\textsc{Argus} across all metrics},
highlighting the ability of GFM-based lateral movement detectors to compete
with GNNs that are specifically trained for this task.
We emphasize that \textsc{Ultra50g}, which underpins \ultralmdpp, was not
trained on any cybersecurity-related data, whereas \textsc{Argus} was trained
on the first 6 days of the OpTC dataset.
This makes the performance of \ultralmdpp\ especially compelling.
Note that the recall at reasonable detection budgets remains rather low for
both detectors.
However, real-world intrusions often comprise several lateral movements, and
detecting some of them can suffice to catch the whole attack.
This principle stands for the OpTC dataset, where
even detecting 5\% of the hundreds of lateral
movement edges is enough to make a detector useful.

As for the ablation study, while \ultralmdpp\ clearly outperforms
\textsc{UltraLMD}, \textbf{most of the
improvement comes from the retrieval component}.
Only adding score refinement leads to almost identical, and in fact slightly
worse performance.
This limited impact of score refinement might result from the short length of
the time window used for the OpTC dataset: with each graph representing a
six-minute window, lateral movement edges are more likely to be scattered
across several graphs.
As a consequence, graph-based score refinement can actually reduce their
anomaly scores as much as those of benign edges.
While this does not refute the general usefulness of score refinement, it does
point to unfavorable settings that can make it less effective.
Still, the performance gain brought by context graph retrieval sustains our
main hypothesis: careful input design can yield superior detection performance
out of the same GFM.

\begin{figure}[t]
	\centering
	\includegraphics[width=\textwidth]{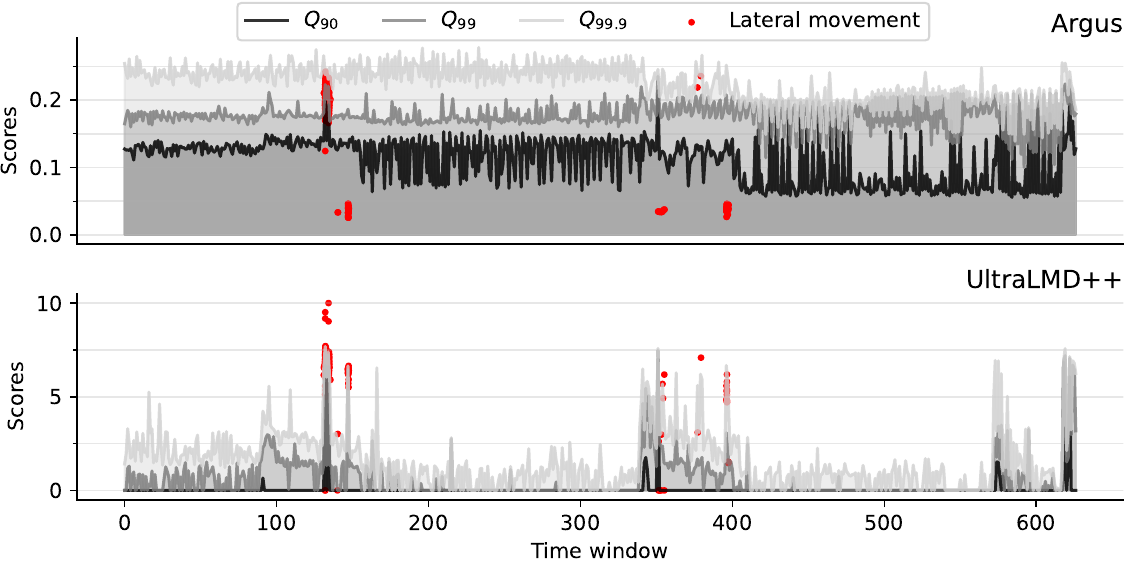}
	\caption{Evolution of the $90^{\mathrm{th}}$, $99^{\mathrm{th}}$, and
		$99.9^{\mathrm{th}}$ percentiles of the distribution of anomaly
		scores over time on OpTC, for \textsc{Argus} and \ultralmdpp.}
	\label{fig:scores_optc}
\end{figure}

\begin{wrapfigure}[12]{R}{0.36\textwidth}
	\centering
	\includegraphics[width=.35\textwidth]{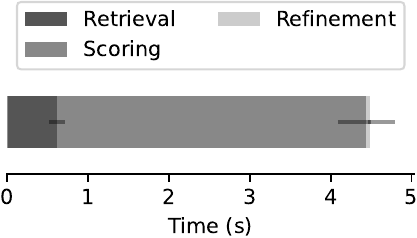}
	\caption{Run time per window (mean and standard deviation) of \ultralmdpp\ 
	on OpTC.}
	\label{fig:times_optc}
\end{wrapfigure}

Regarding score dynamics, Figure~\ref{fig:scores_optc} shows the evolution of
the distribution of
anomaly scores ($90^{\mathrm{th}}$, $99^{\mathrm{th}}$, and
$99.9^{\mathrm{th}}$ percentiles) over time, as well as the scores of lateral
movement edges, for \textsc{Argus} and \ultralmdpp.
Due to the limited duration (three days) of the OpTC test set, no conclusion
can be drawn regarding each detector's propensity to concept drift from
this dataset.
However, we observe that \textsc{Argus} performs better on the first attack
(around window 130) than on the second (windows 350-400), while \ultralmdpp\ 
predicts relatively high anomaly scores for lateral movement edges in both
attacks.
This further demonstrates the superior reliability of \ultralmdpp\ on the OpTC
dataset.

Finally, Figure~\ref{fig:times_optc} shows the average run time of \ultralmdpp\ 
for a single time window on OpTC.
With only a few seconds of computation for a six-minute window,
\textbf{\ultralmdpp\ is efficient enough to run in real time}.
Note that the anomaly scoring step makes up for most of the computational cost,
which could thus be significantly reduced with more powerful GPUs.

\subsection{Results on LANL}
\label{sec:experiments:lanl}

Similarly to our experiments on OpTC, we also run \textsc{Argus} with the same
hyperparameters as the authors on the LANL dataset.
As for \ultralmdpp, we set the number $K$ of past graphs to include in the
short-term context to 10.
Note that many lateral movement edges are repeated several times in the LANL
dataset (i.e., the red team repeated the same lateral movement at different
timestamps).
Since detecting a single occurrence of each lateral movement edge can be
considered sufficient, we also compute the performance metrics after removing
duplicates.
Specifically, for each evaluated detector, we only keep the highest-scoring
instance of each unique lateral movement edge.
The results of our experiments with and without this deduplication step are
shown in Table~\ref{tab:res_lanl}.

{\setlength{\tabcolsep}{.3em}
\begin{table}[t]
	\caption{Results of our experiments on LANL, with and without duplicate
	lateral movement edges.
	}
	\centering
	\begin{tabular}{llrrrrr}
		\toprule
		\textbf{Duplicates}
		& \textbf{Algorithm}
		& \textbf{AUC} & \textbf{AP}
		& \textbf{Rec@}$\boldsymbol{3}$
		& \textbf{Rec@}$\boldsymbol{5}$
		& \textbf{Rec@}$\boldsymbol{10}$\\
		\midrule
		\multirow{5}*{yes}
		& \textsc{Argus}~\cite{xu2024understanding}
		& \textbf{.9749} & \textbf{.0065} & .0100 & .0260 & .0900 \\
		\cline{2-7}\noalign{\vspace{2pt}}
		& \textsc{UltraLMD}~\cite{larroche2024inductive}
		& .7645 & .0020 & .0260 & .0400 & .0720 \\
		& \textsc{UltraLMD} + retrieval
		& .8969 & .0040 & .0300 & .0480 & .0860 \\
		& \textsc{UltraLMD} + refinement
		& .7606 & .0026 & \textbf{.0360} & .0460 & .0820 \\
		\cline{2-7}\noalign{\vspace{2pt}}
		& \ultralmdpp
		& .8938 & .0056 & .0340 & \textbf{.0560} & \textbf{.1040} \\
		\bottomrule
		\multirow{5}*{no}
		& \textsc{Argus}~\cite{xu2024understanding}
		& .9821 & .0056 & .0166 & .0364 & .1126 \\
		\cline{2-7}\noalign{\vspace{2pt}}
		& \textsc{UltraLMD}~\cite{larroche2024inductive}
		& .9394 & .0034 & .0430 & .0662 & .1192 \\
		& \textsc{UltraLMD} + retrieval
		& \textbf{.9920} & .0063 & .0497 & .0795 & .1424 \\
		& \textsc{UltraLMD} + refinement
		& .9328 & .0042 & .0596 & .0762 & .1358 \\
		\cline{2-7}\noalign{\vspace{2pt}}
		& \ultralmdpp
		& .9868 & \textbf{.0088} & \textbf{.0629}
		& \textbf{.0927} & \textbf{.1788} \\
		\bottomrule
	\end{tabular}
	\label{tab:res_lanl}
\end{table}
}

Starting with the results \textbf{without deduplication},
\textbf{\ultralmdpp\ performs best in terms of recall at fixed detection
budgets}, while
\textsc{Argus} performs best in terms of AUC and AP.
In contrast, \textbf{\ultralmdpp\ outperforms \textsc{Argus}
across all metrics when
removing duplicate lateral movement edges}.
The reason for this discrepancy is that we designed our detector to treat edges
that appear in both the short and long-term context graphs as normal.
As a consequence, even though lateral movement edges generally get a high
anomaly score on their first occurrence, they are then included in the context
graphs when they reoccur.
However, since detecting a lateral movement edge on its first occurrence is
arguably good enough, this is not a major flaw.
Besides, even without deduplication, \ultralmdpp\ outperforms \textsc{Argus}
with respect to the recall at fixed detection budgets, which more closely
mirrors the constraints of real-world deployments.
\ultralmdpp\ can thus be considered more effective than \textsc{Argus}.
Again, we emphasize that this level of performance is reached using a GFM
that was not trained for lateral movement detection.

The \textbf{respective contributions} of context graph retrieval and anomaly
score refinement to detection performance are \textbf{more evenly distributed}
than in the OpTC experiment.
Adding either one of these components increases almost all metrics, and adding
both is even better, allowing \ultralmdpp\ to consistently outperform
\textsc{Argus} (whereas \textsc{UltraLMD} does worse than \textsc{Argus} by
several metrics).
Overall, this demonstrates the usefulness of the two improvements we
bring to the original \textsc{UltraLMD} detector, further confirming that
adequately querying a pre-trained GFM has a significantly positive impact.

\begin{figure}[t]
	\centering
	\includegraphics[width=\textwidth]{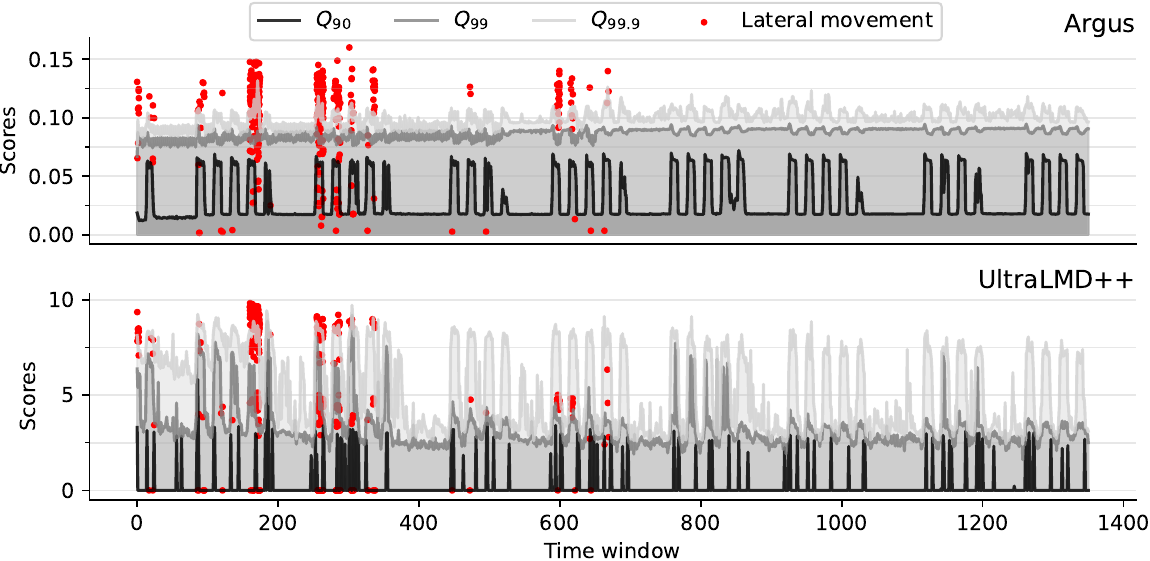}
	\caption{Evolution of the $90^{\mathrm{th}}$, $99^{\mathrm{th}}$, and
		$99.9^{\mathrm{th}}$ percentiles of the distribution of anomaly
		scores over time on LANL, for \textsc{Argus} and \ultralmdpp.}
	\label{fig:scores_lanl}
\end{figure}

\begin{wrapfigure}[12]{R}{0.36\textwidth}
	\centering
	\includegraphics[width=.35\textwidth]{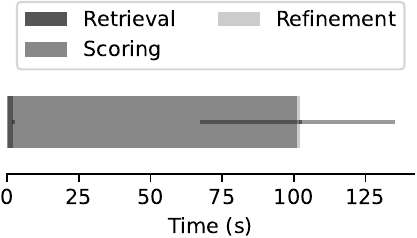}
	\caption{Run time per window (mean and standard deviation) of \ultralmdpp\ 
	on LANL.}
	\label{fig:times_lanl}
\end{wrapfigure}

The evolution of the distribution of anomaly scores over time and the scores
of lateral movement edges for \textsc{Argus} and \ultralmdpp\ are displayed
in Figure~\ref{fig:scores_lanl}.
\textbf{The distribution remains rather stable for \ultralmdpp}, except for a
downward trend in the first 400 windows.
This trend is expected: as more and more past graphs become available, the
context graphs used for anomaly scoring become denser and more similar to the
current graph, thus the new edges appear less and less anomalous.
In contrast, \textbf{evidence of drift can be observed for \textsc{Argus}}.
In particular, the $99^{\mathrm{th}}$ and $99.9^{\mathrm{th}}$ percentiles of
the distribution exhibit a slight upward trend, suggesting that the model
becomes less and less well-adjusted to the data distribution as the new
graphs start becoming less similar to those in the training set.
While this problem could typically be addressed by retraining the model on
more recent data, the need for frequent retraining is impractical in
real-world settings and can hinder the deployment of ML-based detectors.
The ability of \ultralmdpp\ to straightforwardly adapt to distribution shifts
is therefore a significant upside.

Finally, Figure~\ref{fig:times_lanl} shows the average run time of
\ultralmdpp\ for a single time window on LANL.
Similarly to OpTC, \textbf{the run time is low enough for \ultralmdpp\ to be
run in real time} (less than two minutes per one-hour window).
This time is also mostly spent on anomaly scoring.
Overall, the ability of \ultralmdpp\ to run in real time on a 16K-host
enterprise network using rather affordable hardware demonstrates the efficiency
of GFMs, which are typically orders of magnitude smaller than LLMs in terms of
parameter count.
This efficiency is another upside of GFMs for cybersecurity-related
applications, which often deal with sensitive data and must therefore be run
on premise.

\section{Discussion}
\label{sec:discussion}

We now discuss the limitations of our work as well as its possible extensions
and implications for future research.

\paragraph{Future work on GFM-based lateral movement detection.}
First of all, \ultralmdpp\ could still be improved in several directions.
In particular, whether \textbf{fine-tuning a pre-trained GFM} could improve
detection performance remains an open question.
In that regard, a noteworthy aspect of \textsc{Argus} is that it was trained
with a specific loss function, which was shown to yield better
results~\cite{xu2024understanding}.
Fine-tuning GFMs with carefully designed loss functions might thus be a lead
worth investigating.
Future work could also focus on the \textbf{retrieval component} of
our detector: first, the similarity function could be improved so that more
relevant context graphs are retrieved.
Secondly, a current limitation of \ultralmdpp\ is that it needs to observe an
entire graph in order to compute its similarity to past graphs and score its
edges.
Malicious edges can thus only be detected once the time window they
fall into is over, which can be problematic in time-critical situations.
A possible solution could be to train a model to predict which graphs are
the most relevant at a given time, so that these graphs can be retrieved in
advance.
This approach is similar to the one used by Gutflaish et
al.~\cite{gutflaish2019temporal} to dynamically rescale anomaly scores in
the context of user behavior analysis.
Finally, \textbf{experimenting with other GFMs}, such as the more recently
introduced TRIX~\cite{zhang2024trix},
could also increase detection performance.

\paragraph{Further use of GFMs in cybersecurity.}
Beyond lateral movement detection, there is significant potential for useful
cybersecurity-related applications of GFMs.
A straightforward way to identify such applications is to consider past
research on using GNNs in this field.
This includes host and network intrusion detection~\cite{bilot2023graph},
malware detection~\cite{bilot2024survey}, and report analysis for cyber
threat intelligence~\cite{fang2021cybereyes}.
The benefits brought by GFMs in such cases could be the following.
First, by making it possible to develop new applications without training any
model, they could \textbf{accelerate innovation}.
In particular, lowering the required ML expertise would allow
more domain experts to build graph ML-based applications tailored to their own
needs.
Secondly, in the specific case of intrusion detection, GNN-based methods
typically require training one model for each monitored network.
This model must also be regularly retrained to avoid concept drift.
In contrast, our experiments show that a single GFM can deliver high detection
performance in several different enterprise networks without any retraining.
GFMs could thus make GNN-based intrusion detection \textbf{more practical}
in real-world settings.
Finally, applications built upon GFMs would also directly \textbf{benefit
from future research} on these models: since no retraining is needed,
substituting a new
GFM into an existing application to study its performance is straightforward.

\paragraph{GFMs and adversarial machine learning.}
Besides the aforementioned opportunities, the adoption of GFMs in cybersecurity
would also come with potential risks.
Specifically, threats related to adversarial machine learning can be
exacerbated in foundation models.
For instance, creating
\textbf{adversarial examples}~\cite{zugner2018adversarial}
against an openly accessible GFM
is easier than against a private, specifically tailored GNN.
Note, however, that crafting an adversarial example against GFM-based detectors
such as \ultralmdpp\ would require additional knowledge besides the model
itself---namely, since the anomaly scores depend on the context graph, the
adversary would also need access to the data used to build this context graph.
Risks related to adversarial examples thus depend on the application.
\textbf{Data poisoning}~\cite{zhang2021backdoor} is another potential threat
that becomes more serious when
relying on pre-trained models: ensuring that GFMs used in
cybersecurity-related applications are trustworthy and were not trained on
poisoned data would be critical.
Finally, if a GFM was trained specifically for cybersecurity applications and
made openly accessible, the confidentiality of its training data against
\textbf{model inversion attacks}~\cite{zhang2022model}
would be a legitimate concern.
While none of these potential threats definitely precludes the use of GFMs in
the cybersecurity domain, studying them remains instrumental to the adoption of
GFM-based applications.

\section{Conclusion}
\label{sec:conclusion}

The emergence of graph foundation models could be a remarkable opportunity
for researchers and practitioners in cybersecurity.
By removing the need to train task-specific models, GFMs shift the focus onto
other aspects of ML application development, such as input
construction and output post-processing.
This focus shift could both make it easier for domain experts to build
applications relying on graph machine learning, and make these applications
more practical in real-world settings.
Through a detailed case study centered around lateral movement detection, we
provide evidence of this potential by demonstrating that a GFM-based detector
can outperform a more traditional approach that requires training a GNN from
scratch.
These initial results call for further research on the use of GFMs for
cybersecurity.
In particular, the numerous applications of GNNs in this domain are
straightforward starting points for follow-up research.
Investigating the effectiveness of GFM-based systems for these various use
cases will in turn enable a more thorough assessment of the impact of GFMs on
the field at large.

\bibliographystyle{splncs04}
\bibliography{references}

\appendix
\section{Graph-Based Score Refinement Algorithm}
\label{sec:graph}

The anomaly score refinement algorithm introduced in
Section~\ref{sec:design:output} is more formally described below
(Algorithm~\ref{alg:refinement}).

\begin{algorithm}[ht]
  \KwData{
    Graph $\mathcal{G}=(\mathcal{V},\mathcal{E},\mathcal{R})$,
    score map $s:\mathcal{E}\rightarrow\mathbb{R}$.
  }
  \KwResult{
    Refined score map $s$.
  }

  \ForEach{$(u,r,v)\in\mathcal{E}$}{
    \(s_{\max}=\max\left\{
      s[(u',r',v')];\,
      (u',r',v')\in\mathcal{E}\setminus\{(u,r,v)\},\,
      \{u,v\}\cap\{u',v'\}\neq\emptyset
    \right\}\)\;
    \If{$s[(u,r,v)]>s_{\max}$}{
      $s[(u,r,v)]=s_{\max}$\;
    }
  }
  \KwRet{$s$}
  \caption{Graph-based anomaly score refinement.
  }
  \label{alg:refinement}
\end{algorithm}

\section{Additional Details on Dataset Preprocessing}
\label{sec:additional}

As mentioned in Section~\ref{sec:experiments:datasets}, we make different
preprocessing choices than the authors of
\textsc{Argus}~\cite{xu2024understanding}.
We present and justify these choices for the OpTC and LANL datasets in
Sections~\ref{sec:additional:optc} and~\ref{sec:additional:lanl},
respectively.

\subsection{OpTC Dataset}
\label{sec:additional:optc}

We use the flow start events in the host logs to reconstruct internal
network flows.
Only flows between internal hosts are included.
This yields flows between IP addresses, and we replace these IP addresses with
hostnames whenever possible.
Specifically, when a flow start event on a host corresponds to a flow
initiated from this host, we can associate the source IP of the flow to the
name of the host which logged the event.
However, since the host logs of only half of the hosts are present in the OpTC
dataset, we cannot resolve all IP addresses.
We thus leave the unresolved IP addresses as is.
Regarding authentication events, we collect all authentication-related
events from all available hosts.
These include both local and remote log-ins, as well as privileges being
granted to new sessions.
Accounts related to Desktop Window Manager and User Mode Driver Framework
system services (named DWM-* and UMFD-*, respectively) are excluded as they
are generated on the fly.
The main difference between our preprocessing and that of the
\textsc{Argus} paper~\cite{xu2024understanding} lies in the labeling of
lateral movement edges.
Xu et al.~\cite{xu2024understanding} use the same labeling strategy as
Paudel and Huang~\cite{paudel2022pikachu}, who label all flows generated by
compromised hosts after any red team event as lateral movements.
This excessively wide definition generates 21,731 lateral movement edges.
To make the evaluation more realistic, we manually labeled flow start events
that could be traced back to actual lateral movements documented in the ground
truth description of red team activity.
With this approach, only 626 edges are labeled as lateral movements.

\subsection{LANL Dataset}
\label{sec:additional:lanl}

We use the \texttt{auth.txt} and \texttt{flows.txt} files of the LANL dataset
as sources for authentication events and internal network flows, respectively.
The \texttt{redteam.txt} file provides precise labels for lateral movement
edges.
Our preprocessing differs from that of Xu et al.~\cite{xu2024understanding} in
two key aspects.
First, Xu et al. only consider log-ons using the NTLM authentication package,
arguing that other events are unrelated to user authentication.
This is factually incorrect, as many authentications are performed using other
authentication packages (such as Kerberos).
In addition, lateral movements in the LANL dataset happen to coincide with
NTLM authentication events.
Excluding all other authentication packages thus makes lateral movement
detection easier, leading to overestimated detection performance.
We correct this mistake by including all log-on events into the dataset.
The second difference is the temporal scope of the evaluation: while Xu et al.
only consider the first 14 days, we include all 58 days.
Since all red team events are located in the first 30 days, restraining the
scope to the first 14 days mechanically reduces the proportion of benign
edges in the dataset.
Our preprocessing thus leads to a harder, more realistic evaluation.

\end{document}